\documentclass[10pt]{iopart}
\usepackage{iopams}
\expandafter\let\csname equation*\endcsname\relax
\expandafter\let\csname endequation*\endcsname\relax
\usepackage{amsmath}
\usepackage{graphicx}
\usepackage{xcolor}
\usepackage{hyperref}
\usepackage{cite}

\begin{document}

\title[]{Analyzing atomic oxygen product evolution in Micro Cavity Plasma Arrays by a combination of a Multi-PMT OES Setup and a 0-D Chemical Model}

\author{Henrik van Impel$^{1}$, David Steuer$^{1}$, Volker Schulz-von der Gathen$^{2}$, Marc Böke$^{2}$ and Judith Golda$^{1}$}

\address{$^{1}$Plasma Interface Physics, Ruhr-University Bochum, D-44801 Bochum, Germany}
\address{$^{2}$Experimental Physics II: Physics of Reactive Plasmas, Ruhr-University Bochum, D-44801 Bochum, Germany}

\ead{Henrik.vanImpel@rub.de}
\vspace{10pt}
\begin{indented}
\item[]April 2025
\end{indented}

\begin{abstract}
Dielectric barrier discharges (DBDs) are widely used in applications such as ozone generation and volatile organic compound treatment, where performance can be enhanced through catalyst integration. A fundamental understanding of reactive species generation is essential for advancing these technologies. However, temporally resolving reactive species production especially during the initial discharges remains a challenge, despite its importance for controlling production rates and energy efficiency. This study examines atomic oxygen production as a model system for reactive species production in a micro-cavity plasma array, a custom surface DBD confined to micrometer-sized cavities. Optical emission spectroscopy was employed to investigate plasma-chemical processes in helium with 0.1–0.25\,\% molecular oxygen admixture at atmospheric pressure. The discharge, powered by a 15\,kHz, 600\,V amplitude triangular voltage, achieved near-complete oxygen dissociation (up to 100\,\%), as determined via helium state-enhanced actinometry (SEA). A novel multi-photomultiplier system enabled precise temporal tracking of atomic oxygen density and dissociation dynamics. To ensure measurement accuracy, a basic 0D chemical model was developed, reinforcing the reliability of the experimental results.

\end{abstract}
\submitto{\PSST}
\maketitle

\ioptwocol
%
%
%
%
%

\section{Introduction}
Reactive species, such as atomic oxygen, play a key role in many plasma processes. Dielectric barrier discharges (DBDs) were initially developed for large-scale ozone production \cite{Siemens_1857}, where achieving a high dissociation degree of molecular oxygen is essential. Beyond ozone generation, the dissociative potential of DBDs enables applications such as the removal or conversion of environmentally harmful gases, including volatile organic compounds (VOCs) \cite{Koutsospyros_2004}. Further applications are surface treatment \cite{Massines_2001} and biological decontamination \cite{Ehlbeck_2011}. The interaction between conversion and surface effects is of the highest interest for understanding and applying plasma catalysis. The synergistic effects of heterogeneous catalysis and plasma can directly increase energy efficiency and selectivity to desired products, minimize byproduct formation, and optimize mineralization rates. This makes the combination of plasma and catalyst a valuable approach to a green future for industrial applications \cite{VanDurme_2008}.\\
The micro cavity plasma array (MCPA) \cite{Dzikowski_2020} is an interesting candidate for the scientific investigation of the synergy between DBDs and catalysts, as well as direct industrial usage due to its simple scalability properties \cite{Park_2005}. It consists of a multitude of µm-sized cavities where a DBD is ignited at atmospheric pressure. \\
Specific interest in this DBD device was awakened due to the measurement of high degrees of dissociation, revealing an almost complete dissociation of admixed molecular oxygen within the cavities \cite{Steuer_2023, Steuer_2024}. In these papers, atomic oxygen was measured with two different approaches that complement each other: helium state enhanced actinometry (SEA) \cite{Steuer_2022}, a passive optical emission spectroscopical approach, and two-photon absorption laser induced fluorescence (TALIF) \cite{Niemi_2005}, an active laser method. The emission-based method SEA revealed the atomic oxygen density within the cavities where the discharge takes place, whereas TALIF measurements monitored the densities above the cavities. \\
These studies focused on measurements in a steady state. However, the initial discharge cycles are unlikely to exhibit such equilibrium conditions, making them particularly relevant for investigation. Understanding how the dissociation degree evolves during these early cycles is crucial for gaining deeper insights into the dynamics of reactive species within the DBD and identifying potential build-up effects. This is especially important in light of observed discrepancies between SEA and TALIF results, which likely arise from significant losses to the reactor walls \cite{Steuer_2024}.\\
A more comprehensive understanding of these processes could be used to optimize the applied voltage waveform or duty cycle to improve both yield and energy efficiency. To trace the dissociation degree over time, especially within the initial discharge cycles, we developed a novel setup including three photomultipliers (PMTs) recording the line emissions relevant to SEA. This setup allows not only the simultaneous measurement of the line emission but also the evaluation of the dissociation degree on a µs-scale. The setup is presented and evaluated in this work.\\
In addition to evaluating the degree of dissociation, SEA yields further insights into the mean electron energy of the electrons. This energy is crucial for analyzing plasma-chemical processes in greater detail. To facilitate this, we developed a basic chemical model that primarily relies on experimentally measured parameters as input. The model validates the experimental results and provides further insights into the production of ozone or oxygen-related metastables as well as a reliable estimate for wall reactions. The wall losses of these confined cavities are especially essential to explain the given results, revealing a great interaction between the plasma and the surrounding surfaces.\\
This paper aims to provide a comprehensive perspective on the production and loss mechanisms of reactive species in the MCPA. Molecular oxygen serves as a simple model system, offering insights applicable to other reactive species of industrial relevance. The paper is structured as follows: first, the experimental setup, general emission patterns, and discharge dynamics are examined, laying the foundation for subsequent analyses. Next, the temporal evolution of the gas temperature is discussed, as it is crucial for both the SEA evaluation and the presented chemical model. Following these preliminary considerations, the SEA results are presented and compared to the model, offering valuable insights into the dynamics of reactive species.
\section{Diagnostics}
\subsection{Helium state enhanced actinometry}
This section briefly outlines the fundamentals of helium state enhanced actinometry (SEA). A more detailed description of the diagnostic method can be found in other publications \cite{Steuer_2022, Steuer_2023}. SEA is based on classical actinometry, which was used for silicon etch processes, where the intensity ratio of two spectral lines is used to determine the fluor density \cite{Coburn_1980}. This approach was further extended to determine e.g. the atomic oxygen density \cite{Walkup_1986,Katsch_2000}. There, one investigated emission line corresponds to a transition of atomic oxygen, while the other line corresponds to a transition of an actinometer gas with known density. In classical actinometry, only the excitation of the upper states by direct electron impact excitation and the de-excitation by spontaneous emission are considered, while other processes are neglected. Typically, the excited state of oxygen O($\mathrm{3p^3P}$) ($\lambda$=844.6\,nm) is observed in combination with the argon state Ar($\mathrm{2p_1}$) ($\lambda$=750.4\,nm),  whereby argon serves as the actinometer gas. \\
SEA extends this model by including the helium state He(3$^3$S) ($\lambda$=706{.}5\,nm) and the dissociative excitation of O$_2$ as proposed in literature \cite{Niemi_2010,Greb_2014,Tsutsumi_2017}. Considering this helium state further enhances the precision of mean electron energy determination and enables time-integrated measurements \cite{Steuer_2022, Steuer_2023}. The intensity ratios $I_{844}/I_{750}$ and $I_{706}/I_{750}$ are formed from the measurements of the three transitions. The measured intensities are proportional to the amount of the emitting species $f$, the photon energy $\nu_\lambda$, the optical branching ratio $a_\lambda$, and a calibration factor that accounts for wavelength-dependent sensitivity $c_\lambda$ of the setup. From this, the following system of equations is used to calculate the dissociation degree $r_O:=[\mathrm{O}]/(2[\mathrm{O}_2])$ and the mean electron energy:
\begin{equation}\label{prop_E} \frac{I_{706}}{I_{750}} = \frac{c_{706}}{c_{750}} \frac{f_{\mathrm{He}}}{f_\mathrm{{Ar}}}\frac{\nu_{706}}{\nu_{750}}\frac{a_{{706}}}{a_{{750}}}\frac{k_{706,d}(\epsilon)}{k_{750,d}(\epsilon)} \end{equation}

\begin{equation}\label{prop_O} \frac{I_{844}}{I_{750}} = \frac{c_{844}}{c_{750}} \frac{f_{{\mathrm{O}}_2}}{f_{\mathrm{Ar}}}\frac{\nu_{844}}{\nu_{750}}\frac{a_{{844}}}{a_{{750}}}\frac{2r_Ok_{844,d}(\epsilon)+k_{844,de}(\epsilon)}{k_{750,d}(\epsilon)} \end{equation}
Solving the first equation (1) yields the mean electron energy, $\epsilon$, which is then substituted into equation (2) to determine the dissociation degree $r_O$. Specifically, the mean electron energy of the electron energy distribution function (EEDF) is obtained using the BOLSIG+ solver, which solves the Boltzmann equation under the two-term approximation \cite{Hagelaar_2005}. The solver generates a lookup table of effective excitation coefficients, $k_\lambda$, as a function of $\epsilon$. This lookup table is then used to derive theoretical line ratios as a function of $\epsilon$ and $r_O$, which are subsequently compared with the measured values. Besides the direct excitation of the species ($d$), the dissociative excitation of molecular oxygen ($de$) is included. Densities can then be calculated directly from the degree of dissociation by applying the ideal gas law, knowing the gas temperature. More details on the methodology and constants used can be found in another publication \cite{Steuer_2022}.

\section{Setup}
\subsection{Plasma source and waveform}
This study examined the MCPA (see figure \ref{fig:SEA_Setup}), a device consisting of several thousand micro cavities. In this device, an individual micro-discharge is ignited in each cavity. These cavities with a diameter of 50-200\,µm are laser cut into a 40\,µm thick nickel foil (63x20\,mm$^2$). They are arranged in a way that four sub-structures are built. The size of the sub-structures is 10x10\,mm$^2$, and the cavity size varies between the sub-structures. Within each sub-structure, the cavity size and distance (100\,µm) are the same. This foil lies on a dielectric foil (ZrO$_2$, relative permittivity) with a thickness of 40\,µm. Beneath the dielectric foil is a magnet (neodymium) that holds the nickel foil in place. A glass cover, positioned 0.5\,mm above the nickel foil, encloses the entire system, ensuring a sealed gas flow while maintaining the visibility of the DBD emission. A more detailed description can be found in \cite{Dzikowski_2020,Dzikowski_2022,Steuer_2023}.\\
During operation, a high-voltage signal is applied to the nickel foil while the magnet is grounded. The outgoing voltage signal of a function generator (Tektronix AFG 3021B) is amplified (Trek PZD700A M/S) before reaching the nickel grid. A 2$\,$slm gas stream is passed over the nickel foil at atmospheric pressure. The gas composition of helium, molecular oxygen, and argon is controlled via three mass flow controllers (Analyt-MTC Series 358, ranges: 2\,slm, 50\,sccm and 1\,sccm).\\
To measure the temporal development of the mean electron energy and the atomic oxygen density in the first excitation cycles, a voltage waveform in burst mode was used. The used signal is a triangular bipolar voltage (15\,kHz), in which a series of excitation cycles (here 20 lasting for 1.3\,ms) and the length of the interruption until the excitation cycles starts again can be specified. The interruption time was set to 18.7\,ms (yields a duty cycle of 7\,\%) here, so that least no significant volume-based memory effect is to be expected after this time. Moreover, the decay time of atomic oxygen in the effluent of the COST microplasma jet is about 180\,µs under comparable admixture ratios \cite{Steuer_2021}. This means that the interruption time is 100 times longer, and atomic oxygen is not expected to survive, hence allowing it to build-up from one burst cycle to the next. However, it has already been shown that the residual charge carriers on the dielectric have a decay time in the range of hundreds of milliseconds \cite{Labenski_2024}. This means that the first discharge is probably not free of dielectric surface-based memory effects. 
\subsection{Diagnostic setup}
\begin{figure}
    \centering
    \includegraphics[width=\linewidth]{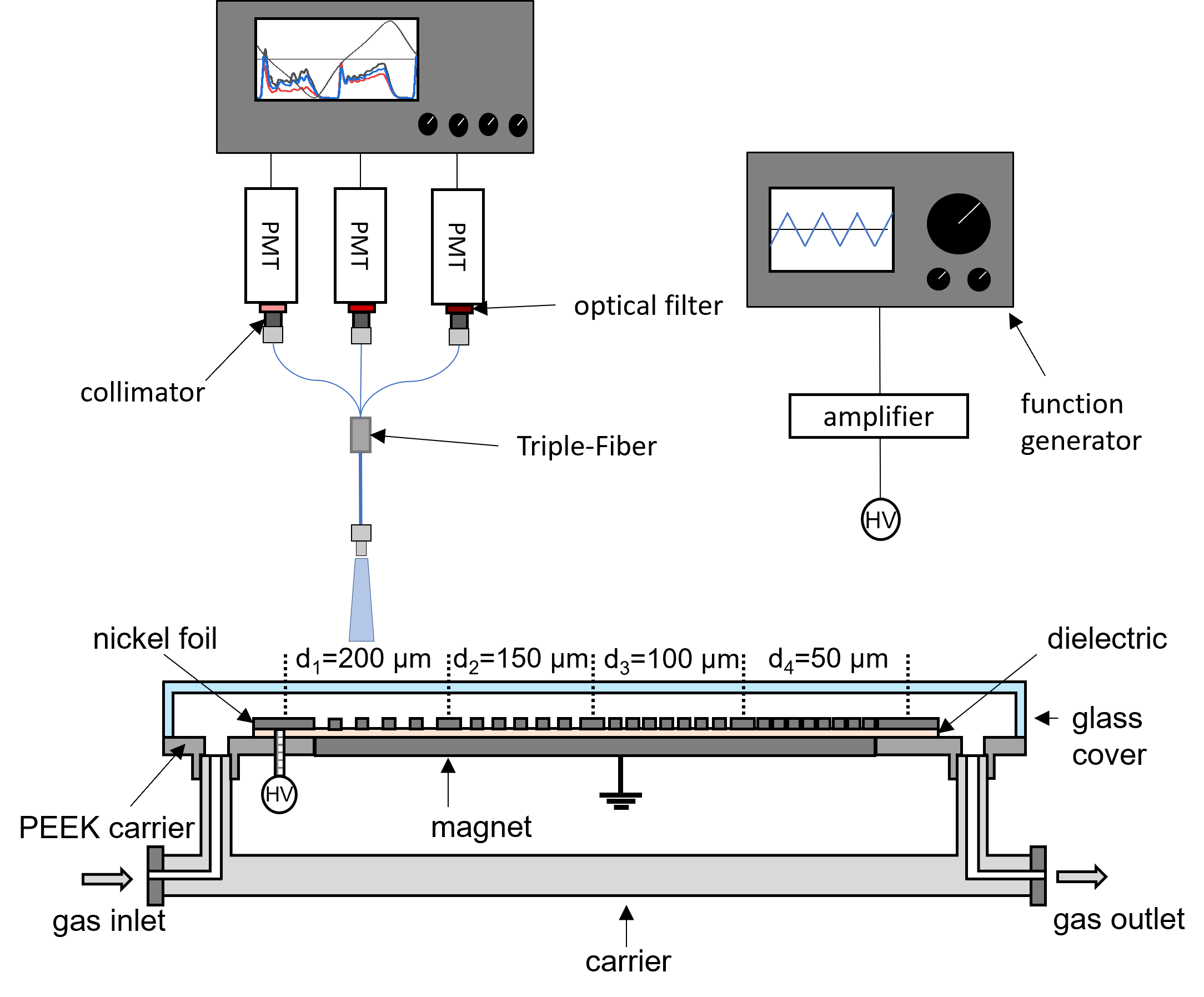}
    \caption{Schematic sketch of the multi-PMT setup and the MCPA.}
    \label{fig:SEA_Setup}
\end{figure}

A novel type of setup consisting of three PMTs (Hamamatsu R13456) was used to measure the temporal development of the atomic oxygen density and the mean electron energy. This setup is shown schematically in figure \ref{fig:SEA_Setup}. A triple fiber (FC3-UVIR400-2-ME) from Avantes is used. The fiber consists of three individual fibers (400\,µm core diameter each), closely confined as one. The merged end collects the emission of the discharge from the reactor at a distance of about 5\,mm, resulting in an emission collection area with a diameter of 1.5\,mm. The three fiber ends are collimated and directed to the PMTs to detect the three relevant SEA lines. The observed wavelengths are separated by three narrowband optical bandpass filters (Thorlabs, see table \ref{tab:PMT}). It must be taken into account that the 751.5\,nm argon line is only partially suppressed by the spectral width of the filter, and therefore the intensity contribution of this transition was estimated. In our discharge, we expect a contribution of less than 10\,\% to the observed line intensity \cite{Steuer_2023} and will neglect any further influence of the additional line intensity on the SEA evaluation. The PMT detecting system was relatively calibrated using a calibration lamp (Ocean Optics DH-3PLUS-BAL-CAL). The calibration factors can be found in table \ref{tab:PMT} revealing the largest sensitivity for the Helium 706\,nm line and the least sensitivity for the Argon 750\,nm line.\\
\begin{table}[h]
\caption{Optical properties of the narrowband optical bandpass filter and the relative calibration factors.}\label{tab:PMT}
\begin{center}
\begin{tabular}{cccc}
\hline
line      & central wavelength & FWHM & $\epsilon_x/\epsilon_{750}$ \\ \hline
He$_{706}$ & 710\,nm   &      10\,nm  &   0.40   \\
Ar$_{750}$ & 750\,nm    & 10\,nm & 1            \\
O$_{844}$  & 840\,nm   & 10\,nm & 0.69             \\ \hline
\end{tabular}
\end{center}
\end{table}The voltage signals from the PMTs are recorded using an oscilloscope (Lecroy 8404M-MS, 4 GHz bandwidth), averaging 512 burst cycles. The oscilloscope is triggered by the function generator to ensure synchronized data acquisition. The integration times or RC times of the PMT systems can be controlled by upstream resistors. These RC times were tuned in such a way that the integration time or temporal resolution is as small as possible while achieving sufficient signal-to-noise ratios. The time constant was set to 1\,µs for all three PMTs.\\ 
From the simultaneously recorded line intensities, ratios are calculated and subsequently used to evaluate the time resolved evolution of atomic oxygen densities and the mean electron energy with SEA. The presented setup allows for high and tunable time resolution.\\
The gas temperature, and therefore the gas density, has a direct effect on the optical branching ratio due to its dependence on the quenching of the excited states. Additionally, it serves as an input parameter for the BOLSIG+ solver. Here we observe the emission of the first negative band of nitrogen ions ($\mathrm{N_2^+}B^2\sum_u^+(v'=0)$ $\rightarrow$ $X^2\sum_g^+(v''=0)$), located at 390\,nm to determine the rotational temperature of this state. With a further fitting routine and the assumption of having a Boltzmann-distributed population of the rotational levels, the rotational temperature can be estimated. The used plane grating spectrometer (PGS-2, Zeiss Jena) was illuminated by an optical fiber collecting the emission of the 200\,µm diameter cavities. The spectrum was recorded with a gateable ICCD camera (Andor iStar DH320T-25U-A3), allowing time resolved measurements. The overall wavelength resolution was about 9\,pm. We assume an equilibrium between the rotational and gas temperatures. The spectrum was fitted with the OES-toolbox \cite{Held_2024} using the data from \cite{luque_1999}. A further description of the method used can be found in the literature \cite{bruggeman_gas_2014,Steuer_2023}.

\section{Results}
The results section is structured to systematically present and analyze the key findings. It starts with an overview of the general emission patterns and discharge dynamics, establishing a foundation for subsequent discussions. This is followed by an analysis of the temporal evolution of the gas temperature, which is crucial for both the SEA evaluation and the chemical model. The SEA results are then presented for different O$_2$ admixtures, and finally, these results are compared and further interpreted using a basic chemical model.
\subsection{General consideration of the emission}
\begin{figure*}
    \centering
    \includegraphics[width=\linewidth]{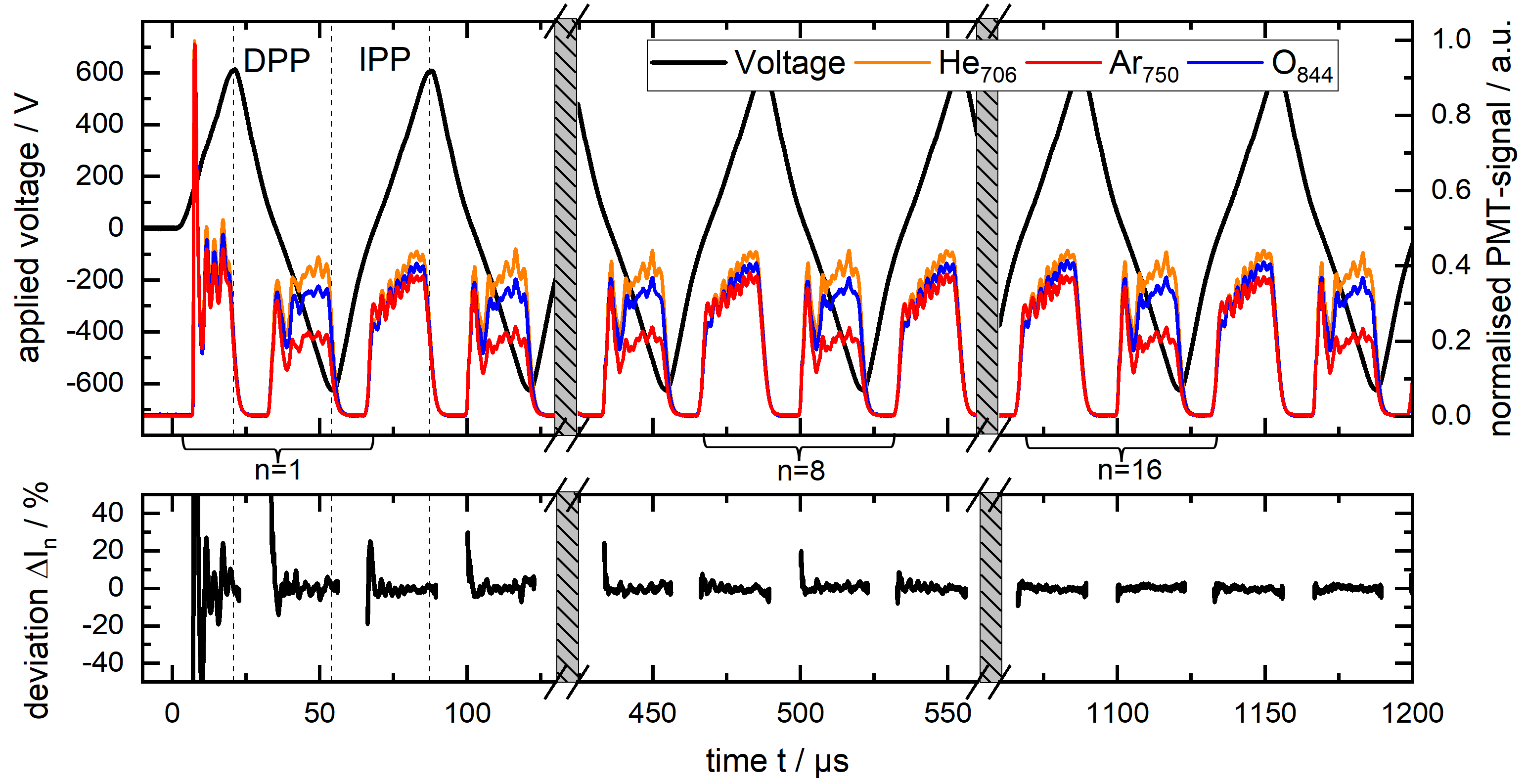}
    \caption{PMT signals at 600\,V applied voltage amplitude (2\,slm He, 1\,sccm Ar, 2\,sccm O$_2$). The lower plot shows the deviation of the He$_{706}$ signal from the reference signal of the last excitation cycle (n=20).}
    \label{fig:Emission}
\end{figure*}

The general discharge dynamics in the MCPA are discussed based on the measured emission patterns. Additionally, the rapid equilibration of the DBD during the series of excitation cycles is examined, supporting the use of averaged measurements for analysis. All measurements were carried out on the sub-array of the MCPA with a cavity diameter of 200\,µm. A representative measurement of the emission pattern is presented in figure \ref{fig:Emission}. The upper part of the figure depicts the temporal evolution of the three PMT signals (He$_{706}$: orange; Ar$_{750}$: red; O$_{844}$: blue) alongside the applied voltage signal (black). The PMT signals are normalized to their respective maximum values. The data was recorded at an applied voltage amplitude of 600\,V with a gas admixture of 2\,slm helium, 1\,sccm argon, and 2\,sccm O$_2$. The amount of actinometer gas (argon) added was chosen to be as low as possible to minimize the influence on the discharge while maintaining a sufficiently high signal strength. To facilitate the comparison of different emission patterns within the applied burst mode, the figure includes two breaks in the time axis, enabling a clearer analysis of the first cycle (n=1), the eighth cycle (n=8), and the sixteenth cycle (n=16). 
\\ 
It should be mentioned at this point that the excitation cycle is divided into two phases (marked by dashed lines): The increasing potential phase (IPP, $\frac{\mathrm{d}U}{\mathrm{d}t}>0$) and the decreasing potential phase (DPP, $\frac{\mathrm{d}U}{\mathrm{d}t}<0$). This choice is reasonable, looking at the emission patterns for different half-phases. The occurring asymmetry of the discharge is mainly induced by the different trajectories of the electron avalanches. In the IPP, the grid is steadily biased more positively. Consequently, the avalanche is accelerated out of the cavity where the electric fields are smaller, whereas in the DPP the avalanche moves toward the bottom of the cavity where large electric fields prevail \cite{Dzikowski_2022,vanImpel_2024}. Further discussion of the asymmetry that occurs can be found in more detail regarding emission patterns, current, electric fields, and surface charges in \cite{Dzikowski_2020,Kreuznacht_2021, Steuer_2023,Labenski_2024,vanImpel_2024, Steuer_2025}.\\
The burst always starts in the IPP and the very first discharge pulse is visible after 7\,µs (n=1). At this very first discharge pulse, a distinct intensity peak is observed, which is at least twice as strong as the emission from subsequent discharges. This behavior can be attributed to the weaker memory effect following the interruption phase, resulting in a higher ignition voltage compared to later discharges. Consequently, the elevated ignition voltage and stronger electric fields provide electrons with greater energy, enhancing atomic excitation. In addition, the increased number of high-energy electrons can generate more ionization and larger avalanches, which can excite more atoms and thus increase the discharge's emission. In subsequent discharges, the memory effect becomes more pronounced, reducing the required effective ignition voltage and consequently lowering the recorded emission intensity. Two key memory effects contribute to this phenomenon. First, the accumulation of metastable species from one discharge pulse to the next increases in the cavity volume. These metastables require less energy for ionization and can generate free electrons through Penning ionization, facilitating ignition.
The second memory effect arises from surface charges accumulating on the dielectric bottom of the cavity. These charges aid ignition, as their binding energy is significantly lower than that of valence-band electrons \cite{Wu_2005}. As a result, they enhance the $\gamma$ coefficient, further promoting ignition and sustaining the discharge process. \\
After the very first discharge pulse, further discharge pulses in the same half-phase follow until about 20\,µs are reached and the voltage drops. This pulsing behavior is typical for a DBD. During the discharge, the dielectric is charged, leading to a shielding of the externally applied electric field until the discharge is extinguished. By further increasing the absolute voltage, the electric field for ignition can build-up again, leading to further discharge pulses. This so-called \textit{pseudo-glow} can also be seen in the following half-phases and is well-known from the literature \cite{Radu_2003,dosoudilova2015}. Usually, the expectation for this discharge type is that with every subsequent discharge, the emission peak becomes smaller as the accumulation of volume charges and metastable species facilitates ignition at lower fields, reducing charge carrier acceleration and multiplication over time. However, in most half-phases, this cannot be resolved since the setup provides an integration time of 1\,µs, averaging about several cavities igniting slightly delayed to each other, leading to a smooth increase of the measured signal. A more precise investigation showing this effect within a single cavity can be found in the literature \cite{Steuer_2025}.\\
In the DPPs and IPPs that follow the initial ignition, ignition occurs even before the voltage crosses zero. This phenomenon is known in the literature as \textit{back discharge} \cite{brandenburg_2017}. The dielectric is charged strongly enough by the discharge of the previous half-phase to form a sufficiently large potential so that the effective ignition voltage within the cavity is reached earlier.\\
In the following, the equilibration of the discharge is analyzed to justify the averaging of signals over multiple burst cycles. Below the PMT signals from Figure \ref{fig:Emission}, the deviation of the measured He$_{706}$ signal $I(t)$ from a reference cycle is shown across the excitation cycles, serving as a representative example for all measured emission lines.. For this purpose, the temporal intensity curve of the last IPP and DPP (20th excitation cycle) $I_{ref}(t)$ from the burst cycle was used as a reference and compared with the curves of the previous half-phases $I_n(t)$, which are indexed here with the index $n$. The calculation of the deviations of the individual cycles weighted with the reference $\Delta I_n$ is as follows:

\begin{equation}
\Delta I_{n}(t)=\frac{I_n(t)-I_{ref}(t)}{I_{ref}(t)}
\end{equation}

The deviations are significantly large during the first IPP and DPP (n=1), exceeding 40\,\% at certain points. This is primarily due to the reduced memory effect, which plays a crucial role in establishing equilibrium. This effect is particularly evident in the initial ignitions of the first two half-phases relative to the reference, where ignition times are significantly shifted, leading to high deviations.
However, these deviations decrease rapidly, dropping below 25\,\% after the first excitation cycle (n$>$1). Notably, the initial discharge in the DPP at the 8th cycle results in large deviations of around 20\,\%, but after the first ignition, the deviations fall below 6\,\%. This indicates that the discharge quickly stabilizes and approaches an equilibrium state after the initial ignition, aligning more closely with the reference cycle. By the 17th excitation cycle, the deviation is already below 3\,\% throughout the whole cycle. This decreasing trend can be attributed to the progressive build-up of the memory effect.\\
In summary, the emission of the discharge, apart from the time of the initial ignition of the half-phases, quickly changes to a stationary state. The deviations from the reference are small and, therefore, rule out a discharge structure that is strongly characterized by stochasticity with the setup measured here. The same analysis of the deviation was also established for the various oxygen admixtures that follow. The deviation is of the same order of magnitude. This means that averaging over several burst cycles can be justified by making general statements about the temporal development of the SEA measurements. However, this does not necessarily apply to the first half-phase of the burst cycle, as a fairly large deviation from the reference can be seen there.
\\
A visual inspection of the emission patterns in figure \ref{fig:Emission} for the He$_{706}$, Ar$_{750}$, and O$_{844}$ signals shows a similar behavior between the observed lines in the different half-phases. However, a detailed evaluation of the relative intensity changes using the SEA approach reveals clear differences in the discharge dynamics related to the atomic oxygen densities and the mean electron energy.

\subsection{Time-resolved gas temperature}
\begin{figure}
    \centering
    \includegraphics[width=\linewidth]{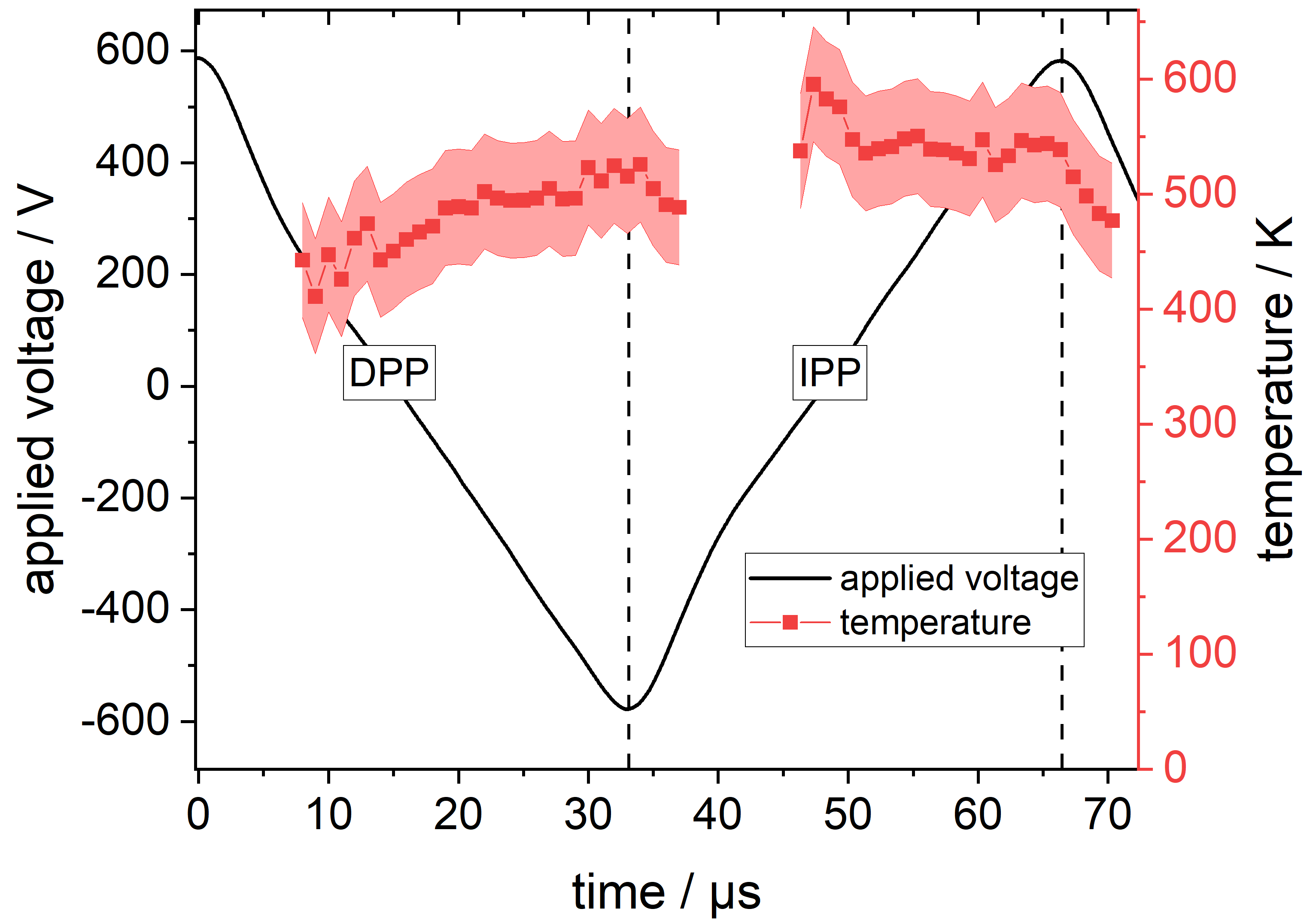}
    \caption{Temporal evolution of the rotational temperature in the DPP and IPP with a 1 µs time resolution at 600 V applied voltage amplitude and 2\,slm He.}
    \label{fig:Temp}
\end{figure}

Since the SEA approach relies on gas temperature to determine quenching constants and absolute atomic oxygen densities, we measured rotational temperatures integrated over the burst mode. The spectrum was integrated for about 100\,s. The measured temperature is about 510\,K at the applied voltage amplitude of 600\,V and smaller than the temperature of 800-850\,K measured in \cite{Steuer_2023}, where a similar reactor was used. The reason for this might be the lower glass cover used here. The lower glass cover has been deliberately chosen to bring the triple fiber as close to the cavities as possible to minimize the averaging effect over the number of cavities. The cover is positioned approximately 0.5\,mm from the nickel foil, whereas the previous one had a distance of 4\,mm. Therefore, the gas velocity is increased by a factor of 8, leading to a more effective cooling of the surfaces and greater exchange of the gas. In addition, the burst mode used here is responsible for a shorter plasma on-time, resulting in less system heating than in continuous operation. However, to further check on the difference between the applied burst and continuous mode and possible temporal effects that can directly influence the correct evaluation, we performed time-resolved measurements in continuous mode. The results are shown in figure \ref{fig:Temp} with a time resolution of 1\,µs. Each measured point was taken with an integration time of 60\,s.\\
The time-resolved measurements in continuous mode reveal that the temperature in the DPP is lower than that in the IPP. Moreover, the trends of the different half-phases are opposite since within the DPP the temperature rises from about 430\,K to 525\,K, while in the IPP the temperature begins to increase from 530\,K to 600\,K and then steadily falls to 540\,K. After each half-phase, marked by the dashed lines, the temperature decreases in both cases due to the drop in absolute voltage, which reduces the electric field supplying energy to the discharge. This is also evident in the pronounced flattening of the emission pattern in figure \ref{fig:Emission} at the end of each half-phase, indicating the discharge's diminishment.\\
The temperature profiles align well with the known asymmetric behavior of the discharge dynamics. The general location of the discharge induces the trend of increasing temperature in the DPP. In that half-phase, the discharges mainly take place within the cavities. Therefore, with each subsequent discharge pulse in the half-phase, the heat of the gas in the cavities is increased and stored until the next pulse heats the gas further. In contrast, the discharges during the IPP are more concentrated above the cavities, where they interact more with the fresh gas flow, leading to rapid equilibration. The temperature peak at the beginning of the IPP might be due to a different discharge mode at the beginning of the half-phase. A transition from a hot streamer-like discharge to an atmospheric pressure glow discharge (APGD) could explain this phenomenon\cite{Steuer_2025}.\\
The general offset between both half-phases can be explained by the location relative to the nickel surface. The overall discharge in the IPP has more of a ring structure and is, therefore, closer to the nickel surface \cite{Dzikowski_2020,Steuer_2023,Steuer_2025}. During operation, the nickel heats up and acts as a heat reservoir. As the method used is emission-based, the heating effect of the gas near the surface is particularly present in the IPP, where the emission is closer to the surface. However, during the DPP, the discharge is more concentrated in the center of the cavity and, therefore, has a larger distance to the heated surface.\\
After analyzing the temporal evolution of the temperature, it is reasonable to use the averaged value of 510\,K, measured in burst mode, for the SEA evaluation.

\subsection{SEA evaluation}

\begin{figure}
    \centering
    \includegraphics[width=\linewidth]{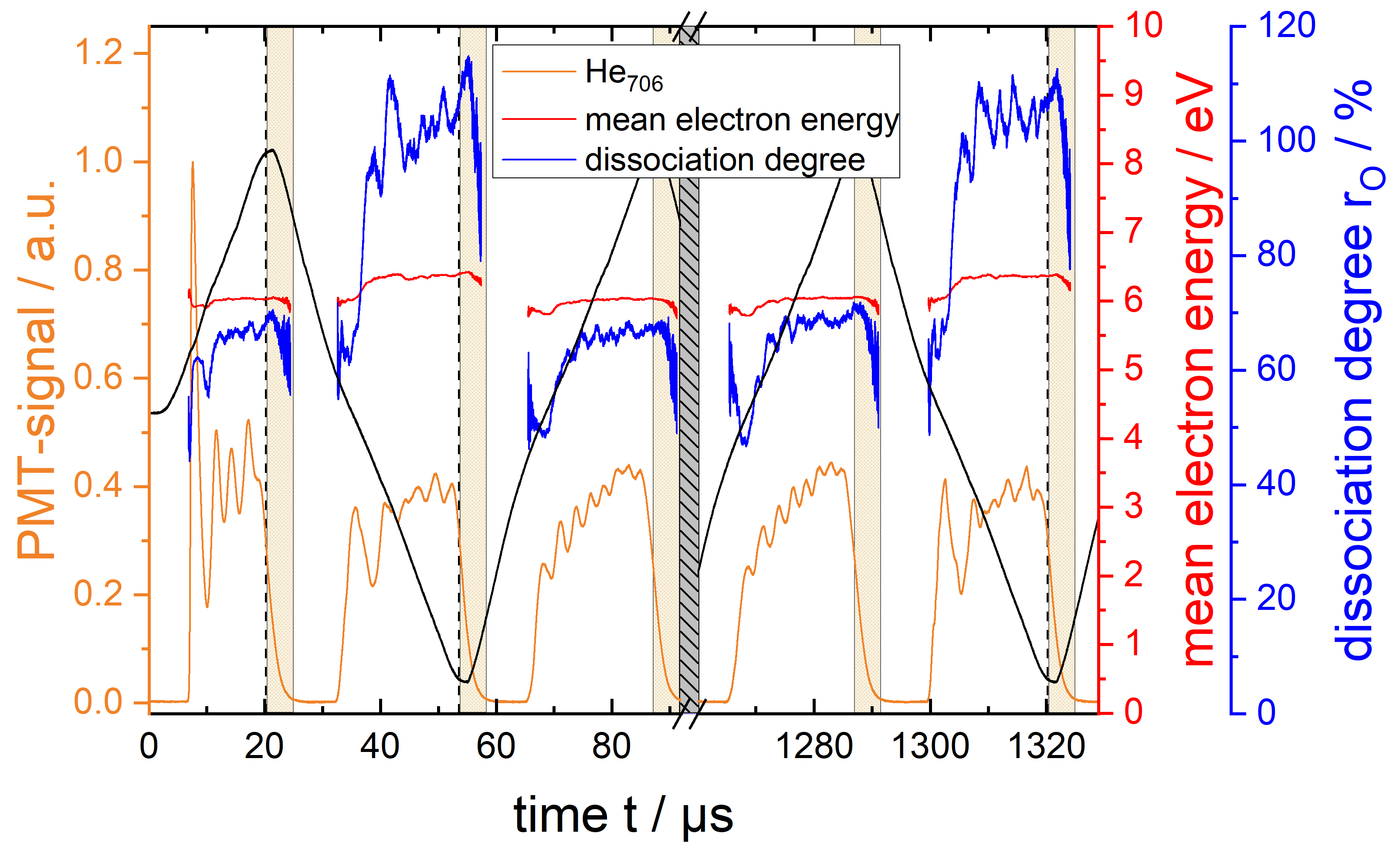}
    \caption{Results of the SEA evaluation with the associated He$_{706}$-PMT signal and the applied voltage (black) with an amplitude of 600 V. The first (n=1) and last discharge cycle (n=20) are shown. Conditions: 2slm He, 1sccm Ar, 2sccm O$_2$.}
    \label{fig:SEA2sccm}
\end{figure}

Following the discussion of the general emission characteristics and discharge temperature, the focus shifts to evaluating the emission signals using SEA. This includes analyzing the temporal evolution of the mean electron energy and the dissociation degree of the admixed oxygen.\\
In figure \ref{fig:SEA2sccm}, the degree of dissociation (blue), the mean electron energy (red), and the normalized PMT signal (orange) of the He$_{706}$ line are plotted for the start (n=1) and the end (n=20) of one burst cycle. The evaluated data belong to measurements with an admixture of 0.1\,\% O$_2$ and an applied voltage amplitude of 600\,V. The evaluation was only performed with sufficient PMT signals, i.e., $>$1\,\% of the respective maximum. The dotted areas at the end of each half-cycle indicate the area where the discharge diminishes and the SEA evaluation appears to be no longer applicable, as the electron driving absolute external electric field stops increasing and the emission drops drastically. As a result, electron impact excitation may no longer be the predominant excitation mechanism in these areas, and the measured signal is therefore determined by cascade excitation or excitation by metastables. However, these types of populations of the excited states under consideration are not taken into account in this actinometry approach.\\ 
The results reveal a degree of dissociation $r_O$ of about 68\,\% in the IPP and about 100\,\% in the DPP after the first ignitions of a half-phase, while the mean electron energy is 6\,eV in the IPP and 6.3\,eV in the DPP. Using the ideal gas law, the atomic oxygen density is therefore around 2.0-2.9$\cdot10^{16}$\,cm$^{-3}$. Here, a temperature of 510\,K was assumed as applied for the SEA evaluation.\\ 
The mean electron energies and atomic oxygen densities have previously been determined using a camera setup with a tunable filter and SEA diagnostics on the same plasma source under identical conditions \cite{Steuer_2023}. These measurements yielded an electron energy of approximately 6.9\,eV and an atomic oxygen density of about 2.2$\cdot10^{16}$\,cm$^{-3}$ for the O$_2$ admixture considered in this study. We have a good agreement with that measurement despite using a completely different detection setup. The small deviations of the mean electron energy can be explained by different influences. First of all, the measured temperature, as well as the applied voltage signal, as described before, were different. In this setup, the lower glass cover was used, increasing the gas velocity by a factor of 8. This resulted in more effective cooling and altered the discharge dynamics while also enhancing the exchange of plasma-treated gas with fresh gas. However, the increased gas velocity does not appear to affect atomic oxygen production. Moreover, in the PMT setup used here, we are averaging several cavities that can ignite at different points in time, and therefore the determined ratios might be spatially more averaged. However, the given setup provides the option to measure simultaneously with a high temporal resolution all three lines, making the measurement in the temporal evolution more accurate.\\ 
Considering the temporal evolution, the characteristic behavior of pulsed plasmas becomes evident in the very first discharge at 8\,µs. Initially, there is a small peak in the mean electron energy of about 6.2\,eV, followed by a high emission peak. This is because a low electron density with a high energy initially generates the charge avalanches. The electron density continues to increase, while the energy is distributed among the electrons and decreases. This high electron density then leads to strong excitation and the subsequent formation of the emission peak.
\\
A key finding of this measurement is that no atomic oxygen density accumulates over the excitation cycles. The degree of dissociation is already at around 65\,\% within the very first discharge half-phase and remains at this level in the following IPP discharges. The same is true for the DPP, where the dissociation degree rapidly reaches 100\,\%. This shows that the very high degrees of dissociation found in \cite{Steuer_2023} are not formed by a build-up over the excitation cycles but that each half-phase dissociates the oxygen almost completely on its own.
\\ 
The asymmetry in the degrees of dissociation and mean electron energies in the different half-phases can be attributed to the electron movement and will be considered separately in the following.
\subsection*{Specifics of the DPP:}
\begin{figure}
    \centering
    \includegraphics[width=\linewidth]{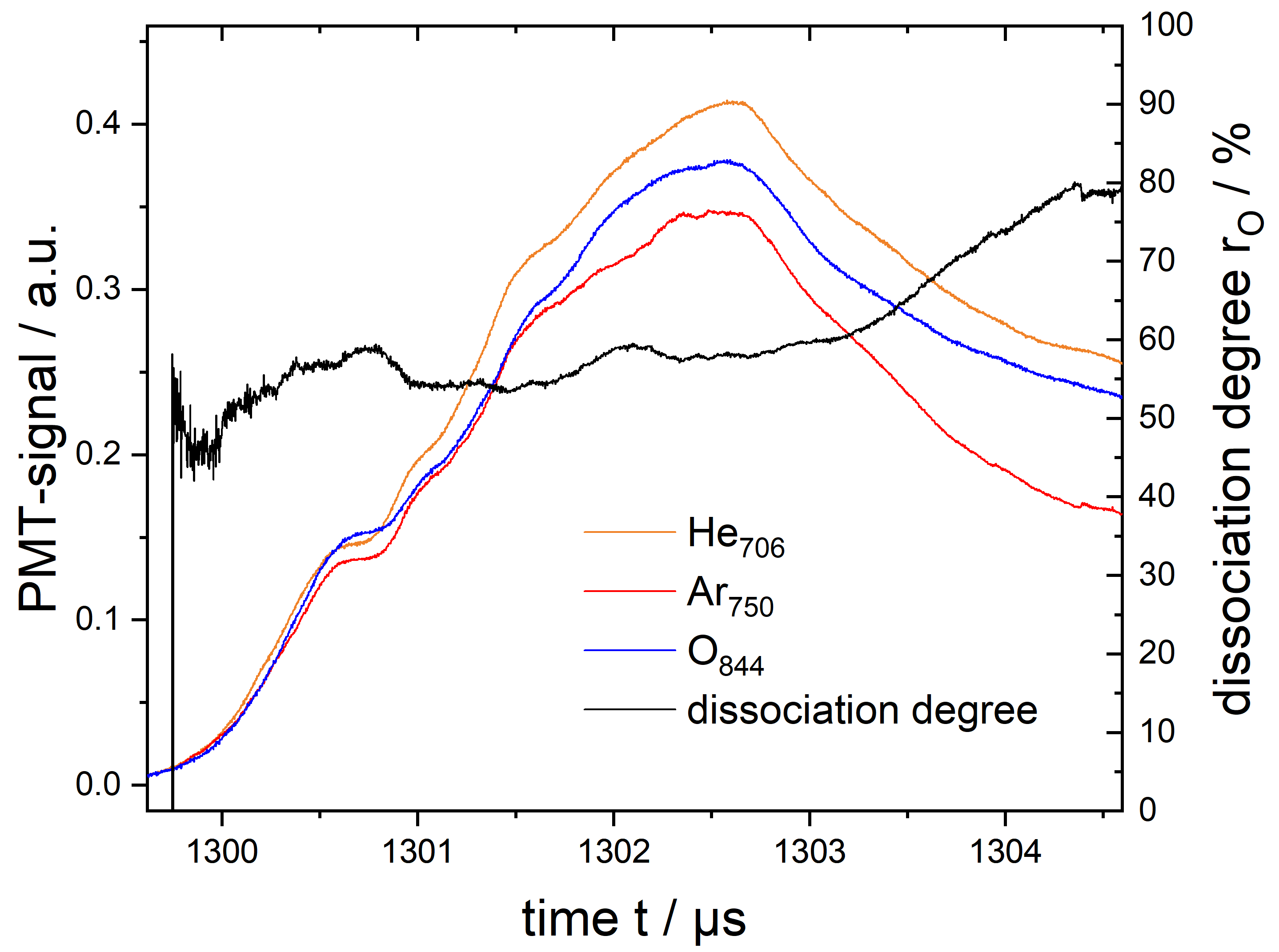}
    \caption{Detailed observation of the increase in dissociation evaluated with SEA with decreasing PMT signals in the last DPP. Conditions: 2\,slm He, 1\,sccm Ar, 2\,sccm O$_2$.}
    \label{fig:Zoom}
\end{figure}

The electron avalanches are accelerated toward the bottom of the cavity, where the high electric fields enable them to gain significant energy. These high energies are then sufficient to generate a high degree of dissociation. In addition, both parameters continue to grow after the first initial ignition of each half-phase. The degree of dissociation increases from 60\,\% to 100\,\% within the DPP from the first to the following discharges in one half-phase. 
This increase in the degree of dissociation can be explained by the fact that atomic oxygen can accumulate within the half-phase from one discharge pulse to the next. Another argument could be the formation of ozone between the discharges, which is easier to dissociate. These build-up effects are enhanced in the DPP by the fact that the discharge and, thus, the built-up species are formed within the cavity and can remain there between discharges. After the first three discharges within the half-phase, however, saturation appears to set in, as the degree of dissociation is beyond 100\,\%. High degrees of dissociation can be artificially elevated by oxidized surfaces that release oxygen during the discharge, effectively increasing the actual oxygen admixture \cite{Steuer_2023}. As a result, values exceeding 100\,\% may occur.\\ 
The increase in the mean electron energy can be directly attributed to the increased degree of dissociation, as less energy of the electrons is transferred to the strongly energy-consuming excitations of the rotational and vibrational levels of the molecular oxygen. In addition, the electronegative atomic oxygen can consume low-energy electrons from the EEDF through attachment processes and thus form negative ions that are not part of the EEDF. Another argument is the increasing voltage, which may allow ignition at locations in the cavity that require higher fields. The electrons are accelerated more strongly by the increased field, and the mean electron energy increases further.\\ 
A notable observation in the DPP is the sharp increase in the degree of dissociation as the PMT signals decrease. This could suggest the unlikely conclusion that dissociation increases when the discharge extinguishes. However, this finding may be influenced by the diagnostic method used and requires further discussion. Since the SEA method is based on OES, it is possible that the emission decreases rapidly in regions where molecular oxygen efficiently quenches excited states. As a result, areas within the discharge with a high dissociation degree may contribute more.
A further reason for the stronger increase is the enhanced influence of metastables and cascade excitations, which are not taken into account in the actinometry model. O$_2$ quenches the excited states with high efficiency. The quenching also includes excitation processes. Thus, it is well conceivable that there are reaction mechanisms that involve, for example, helium metastables and thus lead to a dissociative excitation of O$_2$. According to equation \ref{prop_O}, the intensity ratio of the O$_{844}$ line to the Ar$_{750}$ line is proportional to the degree of dissociation.
This ratio is artificially increased by an uneven occupation of the upper states due to cascade excitation or excitation by metastables and leads to an increased evaluated degree of dissociation. This becomes clear in the PMT signals of the initial ignition of the last DPP, as shown in figure \ref{fig:Zoom}. Here, the three required different emission PMT-signals for the SEA evaluation (orange, red and blue) with the dissociation degree (black) are represented. From about 1303\,µs, the Ar$_{750}$ signal drops faster than the other two signals and thus leads to an increase in the degree of dissociation. The same behavior can also be seen in the dotted areas at the end of the DPPs in figure \ref{fig:SEA2sccm}.\\ 
\subsection*{Specifics of the IPP:}
In the IPP, the electrons are accelerated out of the cavity into the areas of decreasing fields. This means that the mean electron energy is lower than in the DPP. Due to the lower electron energy, the degree of dissociation during initial ignition is also lower than in the DPP.\\ 
In contrast to the DPP, no further accumulation of atomic oxygen density can be seen in the IPP after the first three ignitions, indicated by the three first intensity peaks. The degree of dissociation remains constant at 65\,\%. Since the IPP tends to ignite above the cavity \cite{Dzikowski_2020, Kreuznacht_2021}, the gas composition in the discharge can mix more effectively with the gas flowing above the cavities. As a result, the admixture of new O$_2$ into the discharge volume is more likely here than in the DPP. This also explains the noticeably weaker increase in the degree of dissociation when the IPP ends and the PMT signal decreases. In the DPP, a high number of metastables and cascade excitations of upper states within the cavity slow down the signal decay of the O$_{844}$ line. In contrast, in the IPP, these elevated states are more effectively quenched due to improved mixing with O$_2$, leading to more efficient energy dissipation. As a result, the quenching energy primarily excites the rotational and vibrational levels of numerous O$_2$ molecules rather than just a few, which subsequently contributes to the generation of excited atomic oxygen through additional processes. This effect reduces the artificial increase in the degree of dissociation to a greater extent.\\ 
Moreover, the rapid saturation of the dissociation degree can be explained by the plasma-wall interaction. It is already known that the discharge shows a certain ring structure during the IPP, while in the DPP the discharge is more concentrated to the center of the cavity \cite{Dzikowski_2020, Steuer_2025}. Therefore, the discharge tends to be close to the nickel surface of the grid where loss processes like adsorption of the atomic oxygen play a major role.\\
\subsection{Admixture variation}
\begin{figure}
    \centering
    \includegraphics[width=\linewidth]{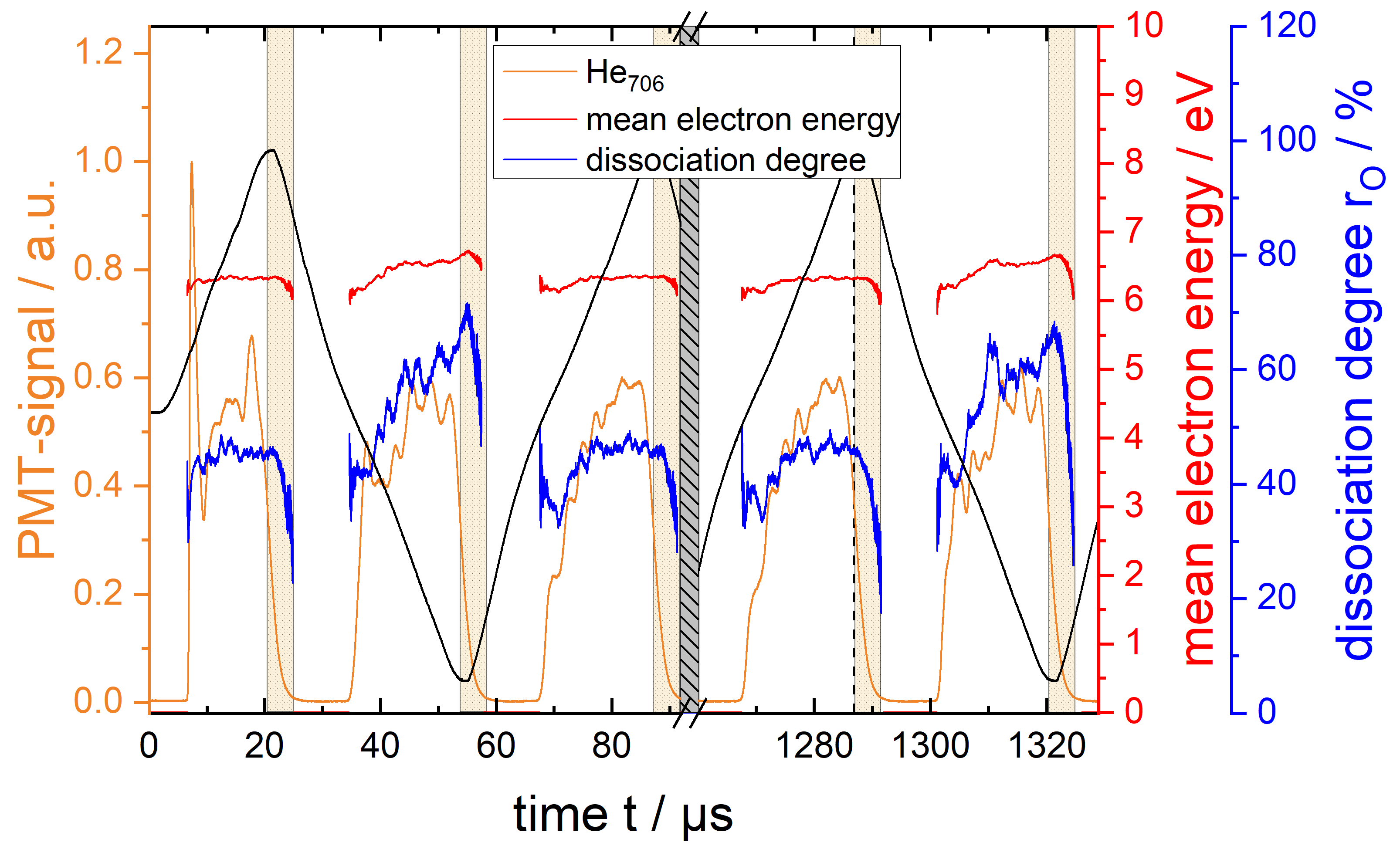}
    \caption{Results of the SEA evaluation with the associated He706-PMT signal and the applied voltage (black) with an amplitude of 600\,V.Conditions: 2\,slm He, 1\,sccm Ar, 5\,sccm O$_2$.}
    \label{fig:SEA5sccm}
\end{figure}

Since a dissociation degree of 100\,\% and thus saturation is observable at an O$_2$ admixture of 0.1\,\%, the admixture was increased to 0.25\,\%. This adjustment allows for evaluating the limits of the discharge's dissociative potential and gaining deeper insights into the dissociation process in unsaturated cases. Increasing the O$_2$ admixture increases the mean electron energy to 6.3\,eV in the IPP and 6.6\,eV in the DPP (see fig. \ref{fig:SEA5sccm}). The degree of dissociation decreases to about 45\,\% in the IPP and 60\,\% in the DPP. The basic development trend does not change compared to the 0.1\,\% admixture. The IPP reveals again a rapid saturation of the dissociation degree, while the DPP shows a more distinct continuous increase of the dissociation degree without saturating. Again, no build-up of atomic oxygen can be seen from one half-phase to the next.\\ 
The general increase in the mean electron energy is caused by the electron attachment of the energetic low electrons by the electronegative oxygen. Despite that, the discharge requires more energy to sustain since quenching is significantly increased, which leads to less ionization and a lower electron density. The lower degree of dissociation is due to the presence of more O$_2$. The discharge cannot expend enough energy to dissociate the additional molecules. This is also evident in the DPP, which does not exhibit clear saturation at higher admixtures, as full dissociation is not nearly reached within that half-phase.

\subsection{Chemical 0-D Model}
\begin{figure}
    \centering
    \includegraphics[width=\linewidth]{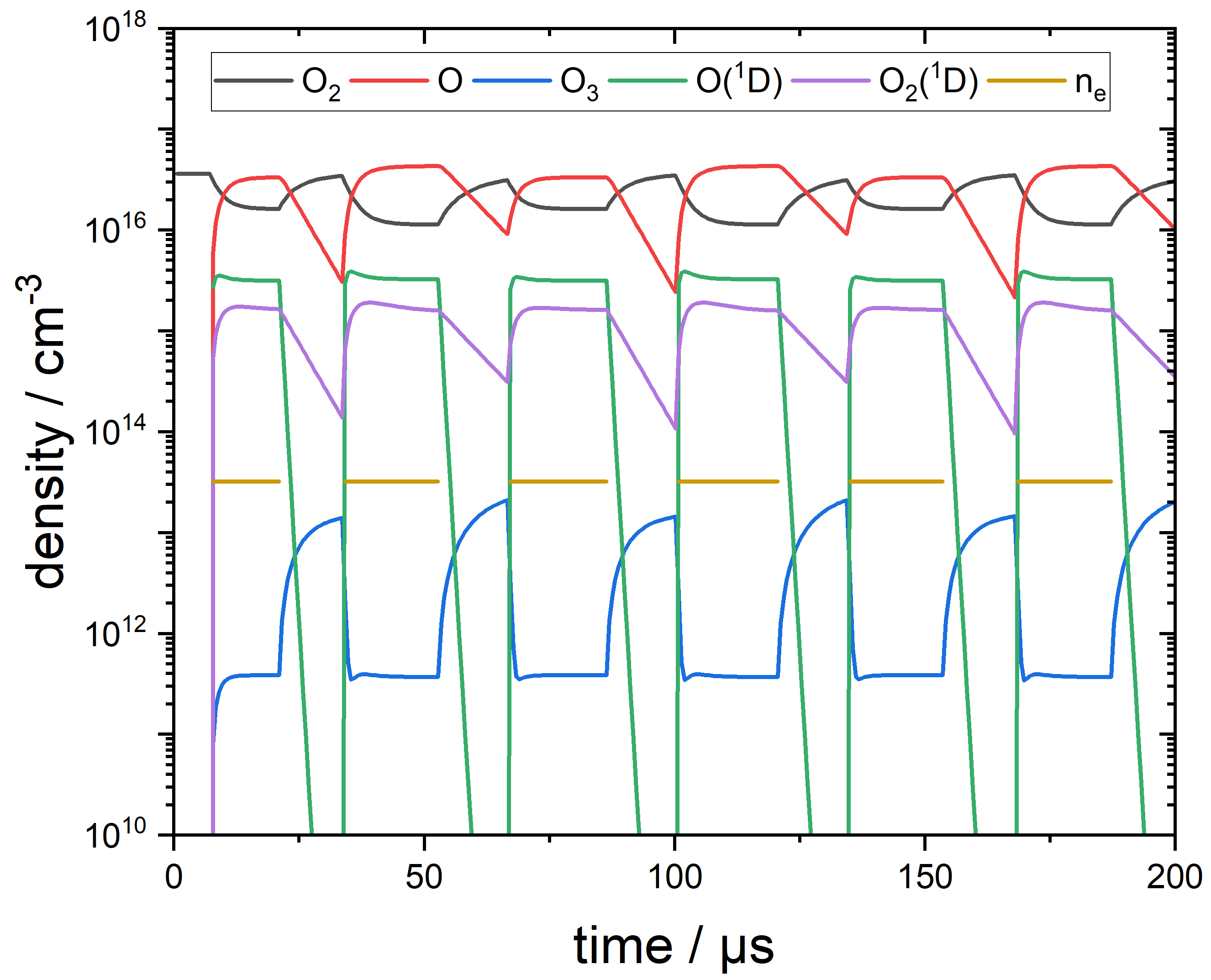}
     \caption{Absolute oxygen densities, as determined by the chemical model, for an O$_2$ admixture of 0.25\,\%. The assumed constant electron density during the plasma on-time is $\mathrm{n}\mathrm{e} = 3.2 \cdot 10^{13}$ cm$^{-3}$.}
    \label{fig:densities}
\end{figure}

\begin{figure}
    \centering
    \includegraphics[width=\linewidth]{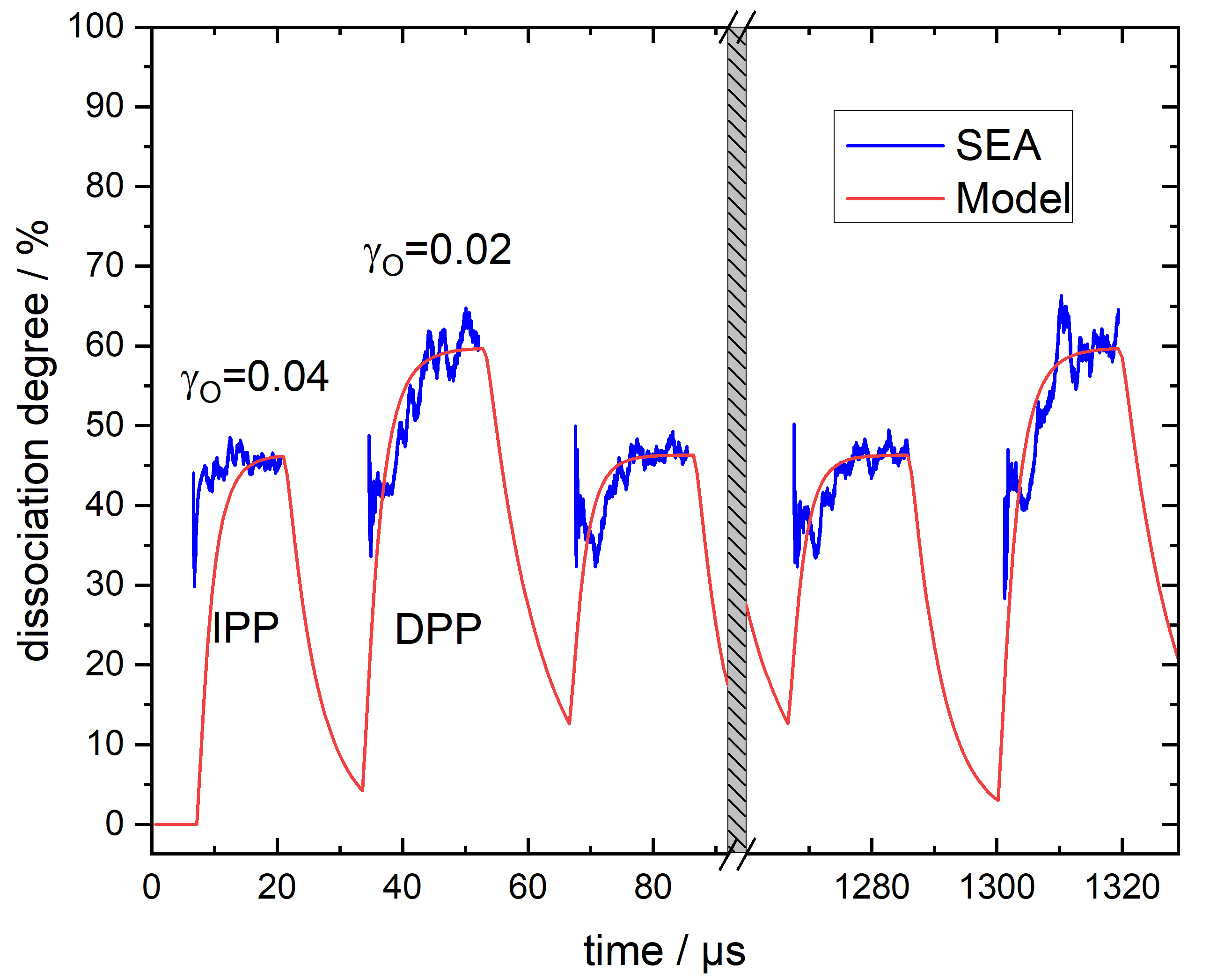}
    \caption{Comparison of the first and last discharge cycle between the chemical model and the SEA measurement for an admixture of 0.25\,\% of O$_2$.}
    \label{fig:comp}
\end{figure}

In order to facilitate a more profound analysis and interpretation of the SEA results, with a particular focus on the origin of the asymmetric behavior observed between the half-phases, a basic 0-D chemical model was developed. This model serves to expand the understanding of the ongoing plasma chemistry within the discharge, thereby providing insights into the gain and loss processes of atomic oxygen. Nevertheless, it allows an estimate of the electron density required to build-up the dissociation degrees evaluated. The whole model is based on the global model from \textit{Murakami et al.} \cite{Murakami_2013}. In this model, we only consider the rate equations \ref{eq:rate} for the helium and oxygen reactions. 

\begin{align}\label{eq:rate}
\frac{d\mathrm{n}_i}{d\mathrm{t}} &= \mathrm{n}_i \sum_{j}^{} \mathrm{k}_{ij}(\mathrm{T}_\mathrm{gas},\mathrm{T}_\mathrm{e}) \mathrm{n}_j \notag \\
&\quad + \mathrm{n}_i \sum_{j}^{} \sum_{m}^{} \mathrm{k}_{ijm}(\mathrm{T}_\mathrm{gas},\mathrm{T}_\mathrm{e}) \mathrm{n}_j \mathrm{n}_m \notag \\
&\quad + \mathrm{R}_\mathrm{ex}(\mathrm{T}_\mathrm{gas})
\end{align}
 The rate equation includes two-body reactions $k_{ij}$, three-body reactions $k_{ijm}$, and external reactions $R_\mathrm{ex}$. We further simplified it by excluding ionized species and only considering the development of the O$_2$, O, O$_3$, O($^1$D), and O$_2$($^1$D) densities. For computing the reaction rates, the gas temperature T$_\mathrm{gas}$ and electron temperature T$_\mathrm{e}$ are required. The model is intentionally simplified by incorporating as many experimentally determined input parameters as possible. Therefore, the model uses the gas temperature from figure \ref{fig:Temp} (IPP: 530\,K; DPP: 500\,K) and the electron temperature from the mean electron energies of the SEA evaluation. The mean value is translated to an electron temperature by multiplying it by 2/3 since the mean value of a Maxwellian distribution is connected like this to the temperature. The transition from the computed EEDF of the BOLSIG+ solver, used in the SEA evaluation to determine the mean electron energy, to a Maxwellian distribution introduces a deviation that should be further discussed. In the model, reactions involving oxygen are primarily considered. However, the largest deviation from a Maxwellian distribution is expected above 20\,eV, where helium, as the carrier gas, is mostly excited. Since the oxygen dissociation occurs around 6\,eV, the impact of using different EEDFs is expected to be minimal. In the model, a constant electron density in the range of 10$^{13}$–10$^{14}$\,cm$^{-3}$ was adjusted during the ignited phases to achieve the best fit to the measured dissociation degree. Typical values for the electron densities in microcavity plasmas from literature lie in the range of 10$^{12}$-10$^{16}$\,cm$^{-3}$ \cite{Eden_2013,Wollny_2011,Kushner_2004}. A more recent study in a similar microcavity discharge device in helium revealed an electron density of about $10^{14}$\,cm$^{-3}$ \cite{Kouadou_2024}.\\
In the model, only the losses of atomic oxygen to the walls are considered, as well as the deexcitation of the metastable states. For this, we estimate that the mean loss time is computed according to \cite{Chantry1987,Booth1991,He2021}:
\begin{equation}\label{eq:loss}
    \tau_{\text{wall}} = \frac{1}{\mathrm{k}_{\text{wall}}} = \frac{\Lambda_0^2}{\mathrm{D}_\mathrm{O}} + \frac{\mathrm{V}}{\mathrm{A}} \cdot \frac{2(2 - \gamma_{\mathrm{O}})}{\bar{\mathrm{v}} \cdot \gamma_{\mathrm{O}}},
\end{equation}
where the diffusion length $\Lambda_0$ was determined for a circular cylinder with the height (50\,µm) and the diameter (200\,µm) of the cavity. V/A describes the volume-to-surface ratio and $\bar{\mathrm{v}}$ the mean velocity of the O-atoms at the respective temperature. The diffusion coefficient was chosen for atomic oxygen to be $\mathrm{D}_\mathrm{O}=1.29\cdot10^{-4}\mathrm{m}^2\mathrm{s}^{-1}$ \cite{Waskoenig_2010}. Moreover, the assumed wall reaction probability for the recombination reaction of atomic oxygen to molecular oxygen on the wall in the IPP is $\gamma_{\mathrm{O}}$=0.04. This sticking coefficient was measured at a nickel surface and is therefore suitable for our nickel foil \cite{Hartunian_1965}. However, for the DPP we assume a smaller effective value $\gamma_{\mathrm{O}}$=0.02 since the discharge is mainly centered in the center of the cavity and therefore has less interaction with the nickel surface.\\
Figure \ref{fig:densities} presents the temporal evolution of the different oxygen species for the first three excitation cycles. The fraction of added oxygen is 0.25\,\% as in the SEA measurement shown in figure \ref{fig:SEA5sccm}. From this measurement, the mean electron energy (IPP: 6.3 eV; DPP: 6.6 eV) as well as the plasma on-time are determined from the emission signal. In the model, this corresponds to the point where the constant electron density (here: 3.2$\cdot10^{13}$\,cm$^{-3}$) is applied. The model reveals a fast increase of atomic oxygen and a decrease of molecular oxygen in the initial discharge. After a few microseconds, the atomic oxygen density saturates, and no further increase is visible. Despite that, a build-up of ozone to a density of $4\cdot10^{11}$\,cm$^{-3}$ can be seen as well as a rapid build-up of the oxygen metastable densities. The O($^1$D)- and O$_2$($^1$D)-state reach a density of $3.2\cdot10^{15}$\,cm$^{-3}$ and $1.8\cdot10^{15}$\,cm$^{-3}$, respectively. After switching off the discharge by setting the electron density to zero, the plasma-generated species are strongly decreasing again except for the ozone density, which is mainly generated by the recombination of molecular with atomic oxygen and increases to $2\cdot10^{13}$\,cm$^{-3}$. The density build-up of ozone saturates after some microseconds due to the high losses of atomic oxygen. Especially the production of ozone is the dominant loss channel of atomic oxygen above the cavity \cite{Steuer_2024}. However, in this case, where we mainly investigate the production within the cavity, the recombination at the nickel surface is the main loss term.\\
The following discharge half-phase (DPP) shows slightly higher plasma-produced reactive species because of the larger electron temperature applied to the model as well as the smaller sticking coefficient, which is decisive for the emerging saturation level. An overall build-up of any species does not occur, as already revealed in the SEA evaluation.\\
 A great agreement was reached in the direct comparison of the dissociation degree provided by the SEA measurements and the model. In particular, the asymmetry in the discharge location, which leads to varying interactions with the nickel surface and an asymmetric development of the dissociation degree, can be reproduced by this basic model simply by effectively adjusting the sticking coefficient $\gamma_{\mathrm{O}}$ for the different half-phases. Figure \ref{fig:comp} presents the direct comparison of the dissociation degree between the model calculations (red) and the SEA measurements (blue) for an O$_2$ admixture of 0.25\,\%. Again, the graph includes a break so that the first and last cycles can be seen within one plot. The model provides further insights into the development of the dissociation degree when the SEA evaluation is not applicable, e.g., the time ranges when the emissions are too weak or not present. This is mainly determined by the mean loss time of atomic oxygen to the walls as calculated from equation \ref{eq:loss}. The value for the IPP is $\tau_{_{\mathrm{wall,IPP}}}$ = 4.6\,µs and for the DPP is $\tau_{\mathrm{wall,DPP}}$ = 7.8\,µs following the different assumed sticking coefficients $\gamma_{\mathrm{O}}$. Moreover, the model provides insights into the densities of other species that cannot be directly measured.\\
Overall, the primary difference between the model and the SEA measurement is that the model does not account for the pulsed nature of the DBD within the half-phases. Since the model assumes a constant electron density during the plasma on-time, the development of the dissociation degree is rather smooth. In the SEA measurement, the pulsed character of the DBD becomes clear in the fluctuating dissociation degree. This indicates that the discharge ignites multiple times within a single half-period, resulting in an effective plasma on-time that is shorter than the assumption made in the model. Therefore, the assumed electron density in the model is likely underestimated. This underestimation in the model helps compensate for the omission of the pulsed nature of the discharge within a single half-period.
 
\section{Conclusion}
The results shown are not only interesting from a plasma chemistry point of view, but also from a diagnostic point of view, as the triple PMT setup in combination with the SEA approach offers great potential for various industrial applications. The setup makes it possible to trace the production of atomic oxygen in the plasma source in real time, which allows a time-resolved measurement of the plasma source with regard to the dissociation degree of O$_2$ and the mean electron energy.\\
In the case of MCPA, the setup in combination with the applied burst mode shows a fast equilibration and a high degree of dissociation within the first discharge cycle, which means that MCPA efficiently produces reactive species. Furthermore, it was found that there is no accumulation from cycle to cycle.
However, a clear accumulating effect is present in the DPP, while in the other half-phase (IPP), a rapid equilibrium builds up. This asymmetric effect can be valuable for plasma catalytic applications since the DPP generates a highly reactive environment in the discharge volume, whereas within the IPP, the rapid equilibration is mostly affected by plasma-wall interaction. Therefore, the placement of different catalytic materials on the confining cavity walls is reasonable for provoking and investigating synergetic effects between plasma and catalyst.\\
The strong interaction with the wall could further be supported by the described basic chemical model provided in this paper. The model is mainly based on measured parameters like gas temperature and mean electron energy. Despite its simplicity, it shows a great agreement with the dissociation degree and, therefore, further checks the consistency of the SEA measurements. The modeled dissociation degree aligns best with an assumed electron density of $3\cdot10^{13}$ cm$^{-3}$, offering a preliminary estimate of the electron density in the investigated DBD reactor. Other densities of interest like ozone and metastable oxygen absolute densities, as well as their temporal evolution, are revealed.
\ack
This work is funded by the Deutsche Forschungsgemeinschaft (DFG, German Research Foundation) via CRC 1316 (project A6).

\section{References}
\bibliographystyle{iopart-num}
\bibliography{references.bib}
\end{document}